# ENHANCED CLUSTER BASED ROUTING PROTOCOL FOR MANETS

Kartheek Srungaram, Dr. MHM Krishna Prasad

Department of Information Technology,
JNTUK-UCEV, Vizianagaram, A.P, India

{kartheek588,krishnaprasad.mhm}@gmail.com

**Abstract.** Mobile ad-hoc networks (MANETs) are a set of self organized wireless mobile nodes that works without any predefined infrastructure. For routing data in MANETs, the routing protocols relay on mobile wireless nodes. In general, any routing protocol performance suffers i) with resource constraints and ii) due to the mobility of the nodes. Due to existing routing challenges in MANETs clustering based protocols suffers frequently with cluster head failure problem, which degrades the cluster stability. This paper proposes, Enhanced CBRP, a schema to improve the cluster stability and in-turn improves the performance of traditional cluster based routing protocol (CBRP), by electing better cluster head using weighted clustering algorithm and considering some crucial routing challenges. Moreover, proposed protocol suggests a secondary cluster head for each cluster, to increase the stability of the cluster and implicitly the network infrastructure in case of sudden failure of cluster head.

## 1 Introduction

Mobile ad-hoc networks (hereinafter, MANETs) are infrastructure less self organizing networks, formed arbitrarily by mobile hosts using wireless links, and the union of which forms a communication network. Routing protocol provides communication beyond the physical wireless range of nodes by relaying on intermediary nodes [1]. Due to the mobility of nodes, the network topology changes frequently, hence, nodes do not familiar with topology of their network. Each node learns about neighbor nodes by listening announcements (using broadcasting of packets) of other nodes [2]. Due to the existing constraints of MANETs, routing should be resource saving. Clustering is one approach to reduce traffic during the routing process, and several authors proposed cluster based routing protocols [1, 3-5]. All the nodes are grouped into clusters and each cluster has one cluster head in addition to many gateways; the cluster head is responsible for its cluster member, whose rebroadcast can cover all nodes in that cluster.

One of the leading protocols in MANET is Cluster Based Routing Protocol (CBRP), proposed by [3]. CBRP is an on demand routing protocol, where nodes are divided into clusters. Initially each node in the network has undecided state. Node starts timer and broadcasts HELLO packet. If it receives a Hello reply from any cluster head then sets it state as cluster member else it makes itself as cluster head but only when it has bidirectional links with one or more neighbor nodes. Otherwise it repeats the procedure with sending HELLO packets.

Clustering is the process that divides the network into interconnected substructures, called clusters. Each cluster has a cluster head and act as a coordinator within the substructure. Each cluster head, in other terms acts as a temporary base station within its zone or cluster and communicates with its peers. Clustering algorithm used in CBRP is a variation of simple õlowest IDö clustering algorithm in which the node with a lowest ID among its neighbors is elected as the Cluster head. Each Node maintains neighbor table and cluster adjacency table which help to has knowledge of network topology.

This paper proposes a new schema for electing cluster heads by considering factors viz., node mobility, power, transmission range and degree of node, and also suggests a secondary cluster head which improves the performance of CBRP, to make the system fault tolerant.

The rest of paper is organized as follows: section 2 presents related work on routing protocols. Section 3 describes the proposed Enhanced CBRP (hereinafter, ECBRP) algorithm and section 4 presents ECBRP evaluation and results analysis. Finally section 5 concludes and directs for future study.

## 2   Related Work

Ref. [6] proposed a new approach to improve cluster stability, namely Smooth and Efficient Re-Clustering (SERC) protocol. In SERC, every cluster head known as primary cluster head (PCH). Each PCH elects secondary cluster head (SCH). When PCH no longer be a cluster head SCH will be cluster head. Since SCH known to all cluster members, the cluster leadership will be transferred smoothly. Each node has four battery power levels when battery power of PCH at critical threshold it transfers its responsibilities to SCH. This approach improves cluster stability and reduces cluster communication overhead.

Ref. [7] proposed Vice Cluster Head on Cluster Based Routing Protocol (VCH-CBRP) by enhancing CBRP specifically designing self-healing of clusters. To enable self-healing they introduced a vice cluster heads concept. In this, after election the cluster head the cluster head sends a notification to each node about this vice cluster head. If the primary cluster head dies for some reason, then vice cluster head advertises itself as cluster head, which reduces the frequency of calling cluster formation algorithm due to mobility/crash of cluster head. So it increases the performance of clustering.

Ref. [1] proposed Cluster Based Trust-aware Routing Protocol (CBTRP), which is a reactive protocol. CBTRP aimed to work in presence of malicious nodes. In CBTRP each node establishes trust among them and maintains it in a trust table. When establishing a route CBTRP ensure that all nodes in the path are trust worthy. If any node detected as malicious, then it will be isolated from the network such that no packets are forwarded through or from it. As mentioned earlier, one of the problems with any cluster based protocol is cluster head failure, which causes frequent execution of cluster formation algorithm. To avoid frequent execution of cluster formation algorithm one solution is secondary cluster head schema. When primary cluster head fails the secondary cluster head takes primary cluster head responsibilities. CBTRP uses local cluster formation algorithm to increase the cluster stability. In local cluster formation algorithm whenever a cluster head detected as malicious the next best trustworthy node in cluster, will be elected as cluster head. This algorithm considers only malicious cluster heads case. But there are other chances to cluster head failure which yields to frequent execution of cluster formation algorithm like cluster head moved away from the cluster and died due to lack enough energy. Therefore, ECBRP handles cluster head failure caused by node mobility and due to lack enough energy.

## 3   Enhanced CBRP

CBRP is a source routing protocol that works based on dynamic source routing. The main idea of CBRP is to divide the network into overlapped or disjoint clusters. Initially each node starts timer and broadcasts HELLO message, which carries neighbor table and cluster adjacency table. Neighbor table contains information about neighbor nodes id, status and link status, where cluster adjacency table has information about adjacent cluster head id and gateway nodes to reach that cluster head.

In CBRP, routing has mainly two phases: Route discovery and Data packets transmission. When source node wants to send data to destination node, source node broadcasts route request (RREQ) packet to cluster heads. After receiving RREQ packet it checks whether destination node is in its cluster or not. If destination node available sends request directly to it else broadcasts to neighbor cluster heads. Before broadcasting each node adds their node id to packet so it may drop packets with its id. When destination node receives RREQ packet it reply to source with route reply (RREP) packet through the nodes which are recorded in RREQ   packet. If source doesnøt receive RREP from destination within some time period, it tries to send RREQ again. When source receives RREP it starts sending data packets.

Many researchers concentrated their studies on CBRP to improve its performance in different factors. The challenges for any routing in MANETs are mobility, resource constraints. Because of these factors the Cluster head may move away from the cluster or die lack of sufficient energy. So, original cluster head election algorithm of CBRP may not gave a better solution to the problem. Hence, this paper presents Enhanced CBRP (ECBRP) algorithm designed to provide a better solution for this problem. Main motivation for this research is to achieve significant impact in the performance of CBRP by improving the cluster stability. This proposed schema improves and develops performance of clustering algorithm than in CBRP.

ECBRP makes use of Weighted Clustering algorithm (WCA) for electing cluster heads [8] adopts a combined weight metric that takes some parameters like ideal node degree, transmission power, mobility and the battery power of the nodes to elect cluster heads. Each node calculates its weight as follows:

$$W_v = w_1 \hat{e}_v + w_2 D_v + w_3 M_v + w_4 P_v .\qquad(1)$$

Parameter $\Delta_v$ represents degree-difference for every node $v$. Degree of the node is nothing but number of neighbors of that node (i.e., nodes within its transmission range), $D_v$ is sum of the distances with all its neighbors. The running average of the speed for every node till current time $T$ gives a measure of mobility and is denoted by $M_v$. $P_v$ implies how much battery power has been consumed.

In equation 1, the first component, $w_1\Delta_v$, helps to avoid MAC layer problems because it is always desirable for a cluster head to handle up to a certain number of nodes. The second component is related to energy consumption because to communicate for a longer distance it requires more power. Hence, it would be better if the sum of distances to all neighbors of a cluster head is less. The third component, mobility of the node, a cluster head having less mobility shows grater improvement on stability of cluster. Last component, $P_v$, is the total (cumulative) time a node act as cluster. Battery drainage will be more for cluster heads comparing to cluster members. WCA also provides the flexibility to adjust the weighting factors according to our network requirements.

Initially each node in undecided state and calculates its weight $Wv$, then broadcasts their ids along with $Wv$ values. When a node received it, it checks for the node with smallest $W_v$ in its list and sets it as its cluster head and makes itself as cluster member. All nodes broadcast their weights along with ids in Hello message. Before every broadcast a node should calculate its weight. Whenever a cluster head received a broadcast message from an undecided node, it will reply with a Hello message immediately. After receiving reply from a cluster head, the undecided node changes its status as cluster member. When a node elected as cluster head it checks for next best node (i.e. node with smallest $W_v$) among cluster members, and broadcasts it as secondary cluster head to cluster members. So when a cluster head dies for some reason, secondary cluster head takes responsibilities of primary cluster head and improves the cluster stability by avoiding frequent execution of cluster formation algorithm. When two cluster heads move next to each other, then one of them will lose its cluster head position. i.e., whenever a cluster head receives a broadcast message from another cluster head, it checks its own weight with that of the other cluster headøs. The one with smaller weight will be cluster head and another one with larger weight makes itself as a cluster member. If any node falls under the transmission range of two cluster heads, then the node joins to the cluster, which having a cluster head with smallest weight ($W_v$). Each cluster is identified by its cluster head id.

Every node maintains data structures to store information about network. Each node has two tables: Neighbor table and Cluster Adjacency table. Neighbor table contains the information of neighbor node id and status (i.e., cluster head or cluster member), whereas cluster adjacency table consists of neighbor cluster id, ids of the gateway node through which the neighboring cluster head could be reached. Note that there may be many gateways to reach neighbor cluster. These tables are updated periodically by Hello messages.

While coming to route discovery and transmission of data, the process is same as CBRP. ECBRP differs with CBRP when a routing failed because of cluster head failure. In CBRP, while transmitting data from source to destination, if route error occurred because of some reason (i.e., the next node in the path may died or moved away from the transmission range of the node which is currently forwarding the packets), the node which found route error will try to salvage the route. Otherwise, it generates route error packet and tells to source. But in ECBR, if route error occurred, the current node first checks whether the next node is cluster head or not. If it is a cluster head, then in the path the cluster head will be replaced by the secondary cluster head of that cluster.

For evaluating the proposed algorithm i.e., ECBRP, is experimented with authors own java simulation framework and performance is compared with original CBRP protocol.

## 4   Evolution and Results Analysis

In this section, the proposed protocol is evaluated using authors own java simulation framework, and compared with CBRP. In this phase of research, authors concentrated mainly in the packet delivery ratio (PDR) of the network to measure the network performance. PDR is defined as the ratio of the number of packets received at the destination to the number of packets sent by the source. In this paper, authors adopted the weighing factors from WCA [8] to compute Eq. 1, as follows: $w_1 = 0.7$, $w_2 = 0.2$, $w_3 = 0.05$ and $w_4 = 0.05$.

Authors assumed that all nodes initially having equal power and power of the node will be reduced when a node transmits (sending its own packets or forwarding) packets and all nodes are having bidirectional links. We have simulated a wireless ad hoc network with 30 nodes in general but nodes may vary up to 100 in some Experiments. The area of simulation is 400mX400m and the transmission range of the node is assumed as 80m. During the simulation the source and destinations were randomly chosen among all the nodes in the network. While coming to node mobility, destination position of each node selected randomly and moves toward the destination with 20mps with 100sec of pause time.

Figure 1, shows the experimental framework, where nodes are distributed randomly and cluster heads and cluster members are represented with blue and black colors respectively, and the dead nodes are represented with red color.

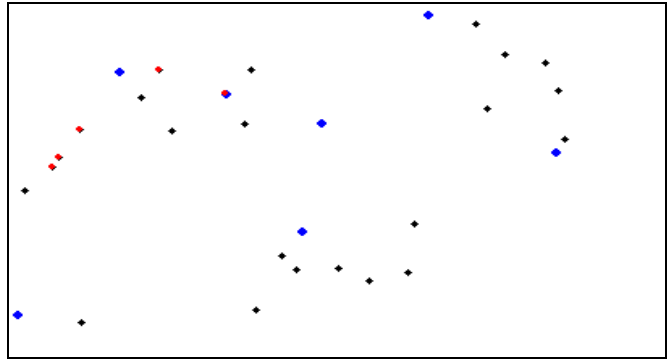

**Fig. 1. Randomly distributed nodes**

As discussed earlier, initially each node calculates its weight using equation 1, and broadcasts it along with their ids. The node having minimum weight will be elected as cluster head and next best one is secondary cluster head, which will be announced by cluster head. Figure 2, shows that a node broadcasting its weight to other nodes which falls within its transmission range.

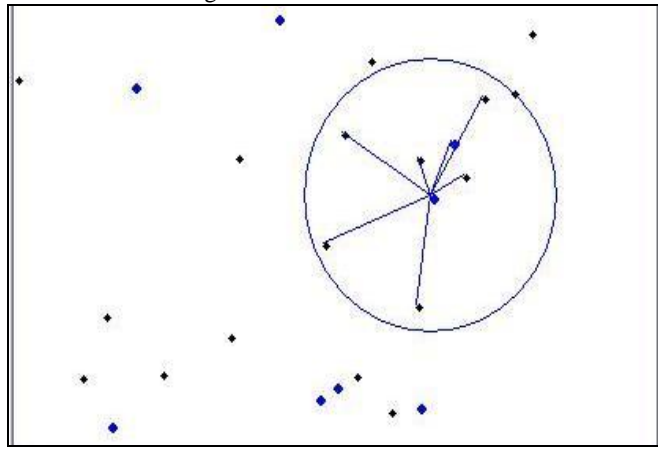

Fig. 2. A node broadcasting its weight (Wv)

The experimental results obtained with ECBRP are compared with the results (PDR) of CBRP, for various numbers of nodes (from 5 to 60), which is presented in Figure 3. For each data point in the result, 5 simulation results were performed, then the average value is computed. From figure 3, one can observe that the Enhanced CBRP performs well than the CBRP at all cases, its due to increase of cluster stability and implicitly reducing the process of cluster reformation.

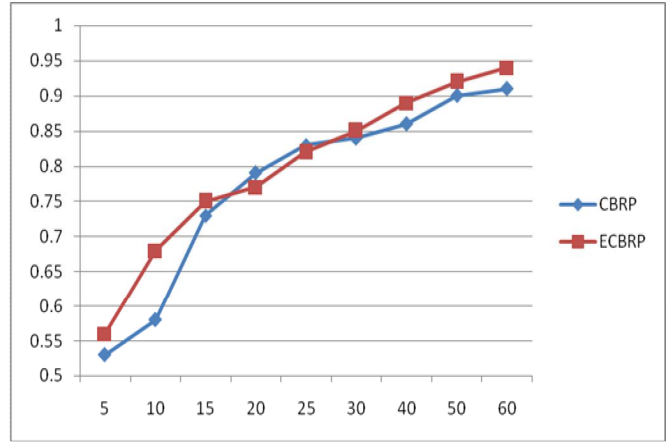

Fig. 3. Results of changing number of nodes to Packet delivery ratio

## 5 Conclusion and Future Work

In this paper, authors proposed an Enhanced Cluster Based Routing Protocol (ECBRP) technique, designed to improve the performance of CBRP. ECBRP makes use of the Weighted Clustering algorithm; ensures the election of best cluster head and improves the cluster stability by replacing cluster heads with the help of secondary cluster heads, leading to rare execution of recreation of clusters. As the performance of network is tightly coupled with the frequency of cluster reorganization, the proposed algorithm helps to reduce the frequency of cluster reorganization and increases the network performance. From the experimental observation it is clear that the Enhanced CBRP performs well, when compared with traditional CBRP.

In general, Intrusion Detection System (IDS) requires support of a good routing protocol. In cluster based IDSs, cluster head runs IDS, which imposes routing and IDS overhead on cluster heads. Due to this phenomenon the cluster head drains quickly. ECBRP helps in a way to decide which node should run the IDS. Authors suggest running IDS on secondary cluster head, which reduces the overhead on cluster head.